\pdfoutput=1
\documentclass[12pt]{article}
\usepackage{tikz,xcolor}
\usepackage{hyperref}

\definecolor{lime}{HTML}{A6CE39}
\DeclareRobustCommand{\orcidicon}{
	\begin{tikzpicture}
	\draw[lime, fill=lime] (0,0) 
	circle [radius=0.16] 
	node[white] {{\fontfamily{qag}\selectfont \tiny ID}};
	\draw[white, fill=white] (-0.0625,0.095) 
	circle [radius=0.007];
	\end{tikzpicture}
	\hspace{-2mm}
}

\foreach \x in {A, ..., Z}{\expandafter\xdef\csname orcid\x\endcsname{\noexpand\href{https://orcid.org/\csname orcidauthor\x\endcsname}
			{\noexpand\orcidicon}}
}
\usepackage[english]{babel}										% Idioma a ser usado
\usepackage[utf8]{inputenc}										% Escrita de caracteres acentuados e cedilhas - 1
\usepackage[T1]{fontenc}										% Escrita de caracteres acentuados + outros detalhes técnicos fundamentais
\usepackage{amsmath,amsfonts,amssymb,amsthm,cancel,siunitx,
calculator,calc,mathtools,empheq,latexsym}
\usepackage{subfig,epsfig,tikz,float}		            % Packages de figuras. 
\usepackage{booktabs,multicol,multirow,tabularx,array}          % Packages para tabela
\title{\normalsize\bf%
\uppercase{3-D Acoustic Trapping With Standing Waves}
}
\author{%
\orcidA Matheus A. S. Pessôa$^{1,2}$ \ and \ \orcidB Antonio Alvaro Ranha Neves$^{2}$
}
\usepackage{geometry}
\newgeometry{vmargin={15mm}, hmargin={20mm,20mm}}   % set the margins

\usepackage[square,sort,comma,numbers]{natbib}

\begin{document}

\date{}

\maketitle

\vspace{-0.5cm}

\begin{center}
{\footnotesize 
*Corresponding author\\
$^1$McGill University, Department of Physics, Montréal, Québec, Canada. \\
$^2$ Centro de Ciências Naturais e Humanas, Universidade Federal do ABC (UFABC), Santo André—São Paulo, 09.210-170, Brazil.\\
E-mails: matheus.pessoa@mail.mcgill.ca, antonio.neves@ufabc.edu.br  
}
\end{center}

% -------------------------------------------------------------------
% Abstract
\bigskip
\noindent
{\small{\bf ABSTRACT.}
In this work, we will describe an experimental setup for a standing--wave ultrasound trap for air microbubbles in oil. We develop a model for the finite acoustic beam using the angular spectrum technique, and reconstruct the pressure field using the General Lorenz-Mie Theory framework, which was validated using a finite elements method (FEM) simulation. Using Stokes' drag law, we were able to obtain the radius of the trapped bubbles and estimate the minimum acoustic force necessary to trap them, which ranged from $3$~nN to $780$~nN. We also present the force profile as a function of distance for different bubble that were trapped experimentally, and show that a standing wave formed by interfering infinite plane waves cannot explain the observed acoustic trapping of bubbles in 3-D. 
}

\baselineskip=\normalbaselineskip
% -------------------------------------------------------------------
                          %display desired
\maketitle
\onecolumn

\section{Introduction}

\quad Acoustic tweezers are devices that enable the non-invasive manipulation of particles on a variety of scales. Thus implementation of such devices in biological studies that involve single-cell manipulation \cite{santos2020raman} and non-organic tissues have grown in the past decade. Examples range from nanometer-scale \cite{cui2019trapping} to micro-scale acoustic traps for solid particles \cite{vyas2019micro} to the manipulation of fluids using ultrasonic transducers \cite{lirette2019ultrasonic}. Biomedical in-vivo applications have been recently experimentally proven, with the manipulation of glass spheres inside the bladder of a living pig \cite{ghanem2020noninvasive}. This development is a milestone in the field of acoustofluidics and acoustic manipulation, with possible future applications to drug delivery systems or kidney stone removal in humans. A biomedical review \cite{guex2021waves}, discusses how current applications with the different approaches in acoustic manipulation (by using surface acoustic waves, bulk acoustic waves, or Faraday waves), device types, and outcomes. With the consolidation and growth of these techniques, biomedical acoustofluidic devices will become even more integrated into society.

For ultrasound manipulation using bulk acoustic waves, the setup in general consists of an ultrasonic transducer being put into contact with a recipient that contains a fluid, and a reflective surface or another transducer directly opposite to the first transducer. As the ultrasonic transducer (or piezoelectric) oscillates, sound waves are generated and transmitted to the fluid. These waves can be either reflected by the opposite surface or interfere constructively and destructively with waves generated by another transducer. In both cases, the resulting scenario are zones of higher and lower pressure within the fluid. This standing wave pattern characterizes a zone in space that has pressure nodes, where there is a resulting acoustic radiation force. Such as light, sound waves also carry momentum and their interaction with matter are feasible when properties such as frequency and the medium in which they propagate are fine-tuned. Microparticles can be trapped and manipulated, depending on parameters such as their size, speed of sound in the medium, and the shape of the incident pressure field, which is crucial for successful trapping. The incident pressure waveform field relies mainly on the geometry of the emitting device, and so does the generated pressure field within the fluid, and the resulting acoustic radiation force responsible for the trapping.

The first experimental observation of the tridimensional negative gradient force that characterizes an acoustic tweezer happened by trapping elastic particles using a single beam \cite{baresch2016observation}. A first characterization of the acoustic radiation force was first reported analytically in 1991 \citep{wu1991acoustical}, when frog eggs were trapped in a potential well. Recent studies \cite{lirette2019ultrasonic} considered the manipulation of droplets on a liquid-liquid interface using CCl$_4$ and water using a finite element method that solved the general fluid form of the acoustic radiation force. Other incident beam shapes have been reported, such as a Bessel beam within or out of the Rayleigh limit, and an evaluation of its analytical expression for the generated force \cite{fan2019trapping}. All 3-D components of the force were obtained in the general case built from momentum conservation arguments and by applying scattering conditions \cite{zhang2018acoustic}.

With acoustic tweezers, it is possible to trap micro-sized bubbles \cite{silva2014acoustic}, which are also objects of study in this work other than the mechanism used for the trap. A remarkable phenomenon that can be achieved by acoustically trapping a micro-sized bubble using a piezoelectric transducer is sonoluminescence \cite{putterman2000sonoluminescence}. Focusing ultrasonic pressure waves on bubbles cause them to rapidly expand and contract. This behavior, driven by the ultrasound waves, is nonlinear and the system is mass-variant as a function of time. When these bubbles collapse they can emit light, whose mechanism is still a matter of consensus among the scientific community. This cavitation process (growth, expansion, and contraction) of bubbles creates an extreme environment that can accelerate chemical reactions \cite{thompson1999sonochemistry} and help in the development of nanostructured materials \cite{suslick1999applications}. Bubbles appear in many industrial applications such as microalgal biomass production \cite{chisti2007biodiesel} and enhancement of membrane processes \cite{cui2003use}. Bubbles also play an essential role in bioreactors, as oxygen gets transferred \emph{via} a rising bubble across the gas-liquid phase \cite{garcia2009bioreactor}. These applications mostly use the fact that bubbles are a gas-liquid interface that can rise when placed in a liquid, serving therefore for the transport process. Bubbles have different shape and dynamics depending on the overall characteristics of the flow in which they are immersed in\cite{kantarci2005bubble}, which are directly linked to fundamental fluid mechanical properties. We make use of some bubble fundamentals to study our standing wave acoustic trap system, to characterize the bubbles in terms of size and shape, and therefore characterize the acoustic radiation force accordingly. This subject has been widely explored in the literature \cite{lee1993acoustic,eller1968force}, but never considering the formalism we use here, starting from pressure and velocity fields and applying GLMT instead of using velocity potentials.

In this work, we present the experimental characterization of the 3-D acoustic trap of air microbubbles. This trap is generated from ultrasonic waves via piezoelectric transducers attached to a recipient containing sunflower oil. The air bubbles were added, by carefully injecting water through a syringe at the bottom of the flask. The water had dissolved air, that released in minor amounts, creating air bubbles that rose and water droplets that fell to the bottom of the flask. The only forces acting on the bubble, when it is trapped, are the acoustic radiation force, and the buoyancy, due to the difference in density between air (bubble interior) and oil (outer medium). We derive the incident field of an ultrasonic piezoelectric transducer using the Generalized Lorenz-Mie Theory (GLMT), and the angular spectrum approach, which results in a realistic description of the experiments.

GLMT is used to describe the acoustic scattering by the bubbles, in terms of the incident, internal, and scattered acoustic fields. With the angular spectrum technique, we reconstruct a more realistic incident beam for the two piezoelectric transducers, as compared to the simple infinite plane wave description, an approach that is validated using finite-element methods.

In Section \ref{methods}, we present the experimental setup in detail and discuss the theoretical framework of the propagation of acoustic waves. In particular, we explore the GLMT formalism, and how to obtain a realistic acoustic beam description, by determining the beam shape coefficients for the transducers using the angular spectrum technique. In Section \ref{sec:results}, we present the experimental results for bubble tracking and acoustical forces, comparing to the case with an infinite plane wave expansion, showing how and why the angular spectrum technique provides an agreement with our experimental conditions. In Section \ref{sec:conclusion}, we present the conclusions of our study and future theoretical-experimental efforts that could be made, using the tools presented in this paper.

\section{Materials and Methods}\label{methods}

\quad The acoustic trap is composed by two piezoelectric transducers that face each other, each at one side of a recipient filled with oil. Using a function generator, we set the piezos to emit sinusoidal waves from each side that trapped microbubbles. After acoustically trapping, the bubbles were released by turning off the function generator, and the dynamics of these bubbles were recorded using a camera. Using an image analysis software, we located the different bubbles, and from their movement calculated their terminal velocities and radius, which allowed the estimation for the acoustic radiation force. 
\subsection{Experimental setup}\label{setup}

\begin{figure*}
\centering

\includegraphics[width=1.0\textwidth]{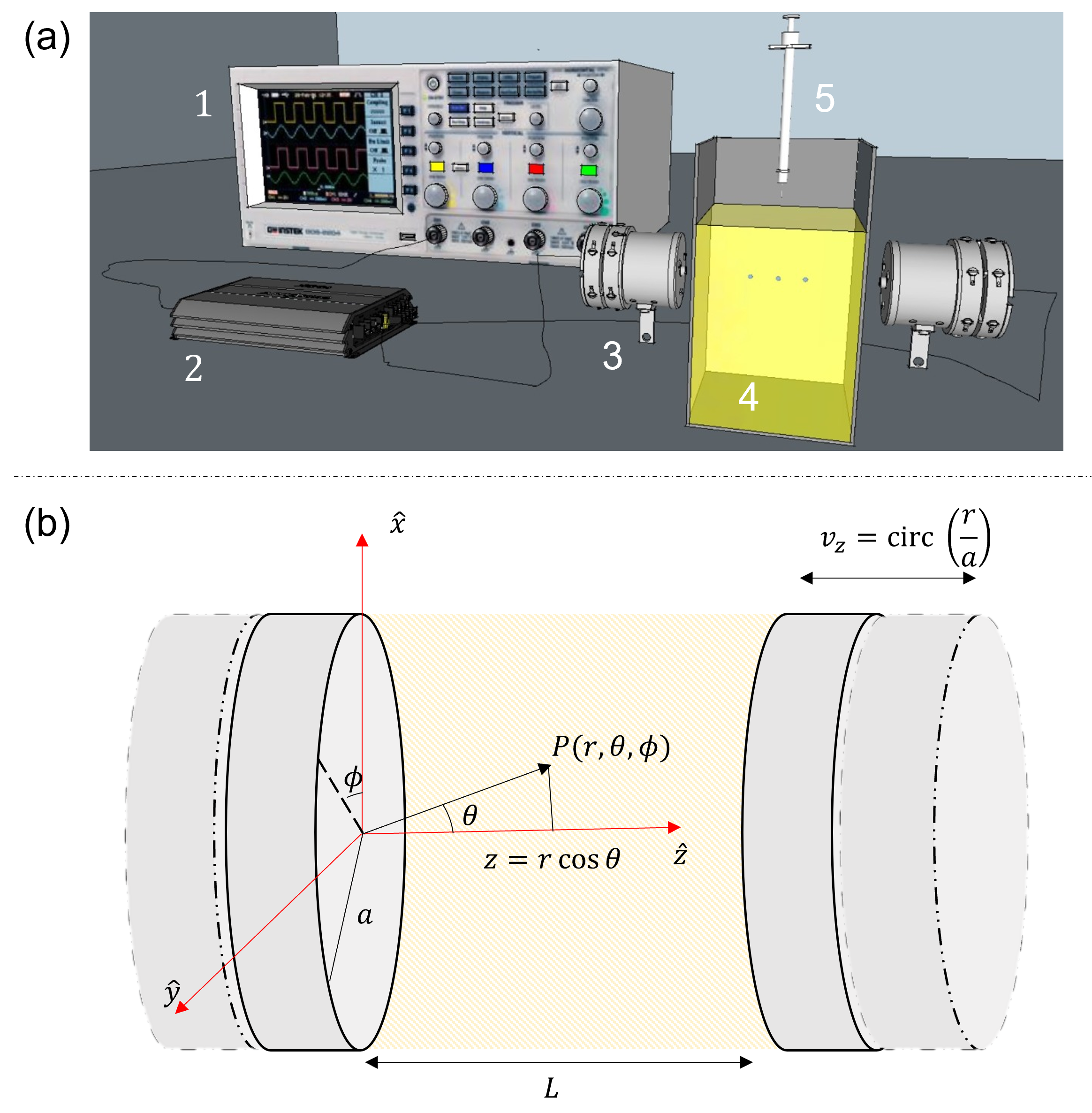}
\caption{Experimental apparatus used for bubble trapping. In 1), the piezoelectric transducers that were used in the ultrasound band from $f=36$kHz to $f=42$kHz, following their maximum operation threshold; in 2), the rectangular flask filled with sunflower oil; in 3), the scaled syringe used for introducing air bubbles in the fluid. The scaling was also used as a reference to measure bubble displacement in the further analysis part. }
\label{fig1}
\end{figure*}
\quad The acoustic trap is composed of a rectangular flask with two centrally aligned piezoelectric transducers on each side. The waves were generated by an arbitrary function generator (Tektronix, AFG3000C) with a $5$V peak to peak voltage sinusoidal wave in the ultrasound band. The generated signal was amplified by a 25W audio amplifier (Taramps, TL-500), powered by a $16$ V  voltage source. The transducers were tightly pressed against the walls of a rectangular flask to produce the acoustic waves by oscillating the walls at ultrasound frequency. This made it possible for the produced waves to transfer momentum to the fluid within the flask creating zones of high and low pressure where the air bubbles were trapped. The air bubbles used were introduced into sunflower oil using a syringe and their displacement was recorded using a \emph{Sony Handycam} 60~fps camera. The schematics for a typical experiment is presented in Figure \ref{fig1}.

For any trapping to occur, both counter-propagating waves must arrive in phase, in such a condition to be achieved an efficient momentum transfer between the acoustic wave and the bubble, enabling the formation of standing acoustic waves inside the flask. For this reason, we performed a frequency sweep between $f=36$~kHz to $f=43$~kHz after the bubbles were introduced into the flask, so we could identify induced displacement or trapping. Trapped bubbles were identified as small bright dots inside of the flask, from light scattered by the external illumination for better visualization. Once trapped, the piezoelectric transducers were switched off, therefore releasing the bubbles. When no longer under the influence of an acoustic force, they would float back to the surface at a given velocity.

As the air bubbles rose to the top of the flask, three forces acts upon it: buoyancy $\boldsymbol{B}=\frac{4}{3}\pi\rho_{oil} g R^3$, weight $\boldsymbol{W}= \frac{4}{3}\pi\rho_{air}g R^3$ and Stoke's drag\cite{lamb1993hydrodynamics}, $\boldsymbol{D}=6\pi \eta R \boldsymbol{v}$ . Considering
$\rho_{air}=1.14$ kg/m$^3$, $\rho_{oil}=988.16$ kg/m$^3$, $\eta \approx 0.04914$ Pa.s for sunflower oil \cite{esteban2012temperature}, at room temperature, it is possible to determine the radius of the bubbles from its terminal velocity using
\begin{equation}
    R= \sqrt{\frac{9}{2}\frac{\eta v}{(\rho_{oil}-\rho_{air})g}}.
    \label{eq:StokesDrag}
\end{equation}
\quad This method was used to characterize the acoustic trap in terms of force, since knowing the bubbles' radii one can infer the buoyancy force and the minimal acoustic force that maintained them in an equilibrium position \cite{crum1971acoustic}. Due to surface tension stresses \cite{chen1999development}, as bubbles move up in a viscous fluid they're susceptible to suffer deformations due to the buoyancy force, reshaping them in toroidal forms for example. These effects are bound to the Reynolds number, expressed in terms of bubble radius, density, and dynamic viscosity of the fluid, $Re=\rho_{air}g^{1/2}R^{3/2}\eta^{-1}$, and the Bond number, expressed in terms of bubble surface tension as well, $Bo=\rho_{air}gR^2\sigma^{-1}$. In this work, the bubble radii obtained did not exceed $Re=10$ nor $Bo=25$, which are in accordance with the characteristics of bubble motion and spherical shape not suffering deformations. These corroborate the observations whilst the experiment was made. 

During experiments, we also noticed that some bubbles would perform a radial movement towards and away from one another, as if there was an apparent attraction and repulsion between them \cite{doinikov1995mutual}. These effects were explained by Bjerknes forces \cite{crum1975bjerknes}, which can also be used for bubble trapping and manipulation \cite{lanoy2015manipulating,yoshida2011experimental}, and to study inter-particle forces \cite{garcia2014experimental}. Although a remarkable phenomenon, in this paper we only analyzed bubbles that were trapped in space, without any rotational or translation movements.

\subsection{Theoretical framework}\label{theory}

\quad In this section, we will first determine the acoustic field produced by a piezoelectric transducer, using the angular spectrum representation. These pressure fields are important, to determine the beam shape coefficients in the GLMT formalism \cite{gouesbet2017generalized}, which is not only used to reconstruct the analytical pressure field but to determine the acoustic forces on any sized scatterer.

As discussed, a single infinite plane--wave propagating in the $z$ direction is unrealistic since it does not consider the finite dimension of the transducer. Also counter--propagating infinite plane--waves, forming a standing wave pattern only trap objects longitudinally but not tangentially, as is observed in some acoustic levitation experiments. To address this need for a more realistic field, we consider the pressure field at distances larger than a few wavelengths from the piezoelectric transducer. 

\subsubsection{Angular spectrum representation}\label{angularspectrum}

\quad We will apply the angular spectrum of plane-waves to determine this pressure field produced by a single piezoelectric transducer. Let the transducer of radius $a$ be located at an input plane $z=0$. The surface velocity, oscillates at the ultrasonic frequency with a constant amplitude, generating a constant pressure amplitude, $P_0$ within the transducer $a$, and zero elsewhere. Let the $2D$ Fourier transform of the pressure field $P(x,y,z)$ be,

\begin{equation}
    \Bar{P}(k_x,k_y;z)=\frac{1}{4\pi^2}\iint P(x,y,z) \mathrm{e}^{-\mathrm{i}(k_x x+k_y y)}\,\mathrm{d}x \mathrm{d}y,
    \label{eq:Fourier}
\end{equation}

and its inverse transform

\begin{equation}
    P(x,y,z)=\iint \Bar{P}(k_x,k_y;z) \mathrm{e}^{\mathrm{i}(k_x x+k_y y)}\,\mathrm{d}k_x \mathrm{d}k_y,
    \label{eq:PressField}
\end{equation}

where $k_x$, $k_y$ are the Fourier coordinates corresponding to the transverse wave vector components. The transform on the plane $z=0$, is related to that at an arbitrary position in $z$ by

\begin{equation}
    \Bar{P}(k_x,k_y;z)=\Bar{P}(k_x,k_y;0) \mathrm{e}^{\mathrm{i} k_z z}.
\end{equation}

Therefore the pressure field can be determined from

\begin{equation}
    P(x,y,z)=\iint \Bar{P}(k_x,k_y;0) \mathrm{e}^{\mathrm{i} k_z z} \mathrm{e}^{\mathrm{i}(k_x x+k_y y)}\,\mathrm{d}k_x \mathrm{d}k_y,
\end{equation}

restricted to the condition where $k_x^2+k_y^2 \leq k^2$. At the plane $z=0$, where the pressure field amplitude is $P_0$ for $r\leq a$, we have,

\begin{equation}
    \Bar{P}(k_x,k_y;0)=\frac{P_0}{4\pi^2}\iint \mathrm{e}^{-\mathrm{i}(k_x x+k_y y)}\,\mathrm{d}x \mathrm{d}y.
\end{equation}

We can rewrite the cartesian components in cylindrical coordinates where $x=\rho \cos\phi$, $y=\rho \sin\phi$, and $k_x=k_\rho\cos\psi$ and $k_y=k_\rho\sin\psi$. Therefore, we can write the 2D Fourier transform of the pressure field as 

\begin{equation}
\begin{split}
    \Bar{P}(k_x,k_y;0)&=\frac{P_0}{4\pi^2}\int_0^a \int_0^{2\pi} e^{-i k_\rho \rho\cos(\phi-\psi)}\,\rho d\rho d\phi\\
    &=\frac{P_0}{2\pi}\frac{a}{k_\rho} J_1(k_\rho a),
\end{split}
\end{equation}
where $J_1(k_\rho a)$ represents the first order of the Bessel function of first kind, calculated in terms of the size of the transducer $a$ and the wave spatial frequency $k_\rho$. From the knowledge of the transform at the plane $z=0$, we can determine the pressure field using Eq.~\ref{eq:PressField}. Now, we can rewrite $P$ in spherical coordinates, where $(x,y,z)=r(\sin\theta\cos\phi,\sin\theta\sin\phi,\cos\theta)$. This yields,

\begin{equation}
    P(r,\theta,\phi)=\frac{P_0 a}{2\pi}\iint J_1(k_\rho a) e^{i k_z r\cos\theta} e^{i k_\rho r\sin\theta\cos(\phi-\psi)}\,\frac{dk_x dk_y}{k_\rho}.
    \label{eq:fullPfield}
\end{equation}

The integral can be simplified by also rewriting the spatial frequencies in spherical coordinates as, $(k_x,k_y,k_z)=k(\sin\eta\cos\psi,\sin\eta\sin\psi,\cos\eta)$, therefore $dk_xdk_y/k_\rho = dk_\rho d\psi=k\cos\eta d\eta d\psi$ where $0\le\eta\le\pi/2$,

\begin{equation}
\begin{split}
    P(r,\theta,\phi)&=\frac{P_0}{2\pi}ka\iint  J_1(ka\sin\eta) e^{i kr\cos\eta\cos\theta} e^{i kr\sin\eta \sin\theta\cos(\phi-\psi)}\cos\eta\,d\eta d\psi\\
    &= P_0 ka\int  J_1(ka\sin\eta) e^{i kr\cos\eta\cos\theta} J_0(kr\sin\eta \sin\theta)\cos\eta\,d\eta.
    \label{eq:FinitePressureField}
\end{split}
\end{equation}

This pressure field will be reconstructed in Section \ref{sec:results}, and compared to other results. We haven't solved the last integral from the lack of a simple closed--form that we are aware of.

\subsubsection{Pressure field between transducers}

In the previous section, we derived an expression for the pressure field at a distance $z$ from our origin. For the standing wave setup, we place the two transducers in such a way that their pressure fields are aligned and counter--propagating. This is done by placing one transducer at $-L/2$, whose fields propagate towards the positive $z$--axis, while the second at $+L/2$, whose fields propagate towards the negative $z$--axis, where $L$ is the distance between the transducers.

This requires that in Eq.~\ref{eq:FinitePressureField}, a phase term $e^{i\frac{kL}{2}\cos\eta}$ when propagating in the positive or negative $z$--direction. For the beam propagating in the positive direction, we replace $z \rightarrow z + L/2$. For the beam propagating in the negative direction, we replace $k_z \rightarrow -k_z$ and $z \rightarrow z - L/2$. Upon combining both contributions from both transducers, the total pressure field can be written as,
\begin{equation}
    P(r,\theta,\phi)=2P_0 ka \int J_1(k a \sin\eta) \cos(k r \cos\theta\cos\eta) \mathrm{e}^{\mathrm{i} kL/2 \cos\eta} J_0(kr\sin\theta\sin\eta)\cos\eta\, d\eta.
    \label{eq:FiniteStatWavePressureField}
\end{equation}

The presence of Bessel functions on the pressure field already provides some insight into how it will behave. Each $n$-th order of the Bessel's functions $J_n (x)$ of the first kind has an oscillatory behavior that decays in amplitude as a function of $x$. In our case, as we are making a finite expansion of the waves, we deal with two situations: the behavior of the pressure along the surface of the transducer, and how the pressure propagates throughout the medium, towards the other transducer. The presence of Bessel's functions can already indicate the finiteness of the solution, and the decaying value of pressure as we analyze the immediate locations around the transducers, as will be discussed in Section \ref{sec:results}.

\subsubsection{GLMT formalism}

The propagation of sound waves is governed by Navier--Stokes equations, with which we can obtain numerical and analytical solutions. The equation can be written as

\begin{equation}
  \nabla ^2{P}=\frac{1}{c^2}\frac{\partial^2 P}{\partial t^2},
  \label{fieldeq}
 \end{equation}
 
for the behavior of the pressure fields involved in our system of study. Eq.~\ref{fieldeq} is the Helmholtz equation, and in our case can be solved using spherical coordinates. In our problem, we can consider two different regions of interest: the inside of the spherical bubbles (with sound speed $c^I$ and $k^I$), and the external region, where we have the incident and scattered waves (with sound speed $c^E$ and $k^E$). The solution for an incident field, for example, can be written in terms of a spherical expansion as
 
 \begin{equation}
     P_{\mathrm{inc}}(r,\theta,\phi)=P_0 \sum_{l=0}^{l_{\mathrm{max}}} \sum_{m=-l}^{l}G_{lm}j_l(k^E r) Y_{lm}(\theta,\phi),
     \label{eq:Pinc}
 \end{equation}
 
where $G_{lm}$ represents the beam shape coefficient, $j_l(k^E r)$ the spherical Bessel's functions relative to the external region of the sphere, and $Y_{lm} (\theta,\phi)$ represents the spherical harmonics. Under the generalized Lorenz--Mie theory formalism (GLMT), the beam shape coefficients can be determined from,

\begin{equation}
    G_{lm}=\frac{1}{P_0 j_l(k r)} \int P(r,\theta,\phi) Y^*_ {lm}(\theta,\phi)d\Omega
    \label{eq:DefG}
\end{equation}

where the pressure field, $P(r,\theta,\phi)$, is that of the standing wave pattern derived before, given by Eq.~\ref{eq:FiniteStatWavePressureField}. Since the pressure field has an integral equation in $\eta$, with the solid angle in Eq.~\ref{eq:DefG}, we have two more in $\theta$ and $\phi$. We can first integrate in $\phi$, using the fact that

\begin{equation}
    \int e^{-im\phi}\,d\phi=2\pi \delta_{m,0},
\end{equation}

where the Kronecker delta, $\delta_{m,0}$, indicates an azimuthal symmetry in our adopted system of coordinates. We, therefore, recover the following beam shape coefficient,
\begin{equation}
\begin{split}
    G_{l,0}&=\frac{4\pi ka}{j_l(k r)} \sqrt{\frac{2l+1}{4\pi}}\int\,d\eta \cos\eta J_1(ka\sin\eta) \mathrm{e}^{\mathrm{i} kL/2 \cos\eta}\\
    &\int\,d\theta \sin\theta P_l^0(\cos\theta) J_0(kr\sin\eta\sin\theta) \cos(k r \cos\theta\cos\eta).
    \end{split}
\end{equation}

The integral in $\theta$, has been previously solved in Ref.~\cite{neves2006analytical,pessoa2020acoustic}, for the special case of $m=0$, resulting in
\begin{equation}
\int_0^{\pi}\,d\theta \sin\theta P_l^0(\cos\theta) J_0(kr\sin\eta\sin\theta) e^{\pm i kr\cos\eta\cos\theta}=2 (\pm i)^l P_l^0(\cos\eta)j_l(kr).
\end{equation}

This yields
\begin{equation}
    G_{l,0}=2 ka \sqrt{\pi (2l+1)} \cos(l \pi/2) \int_0^{\pi/2}\,d\eta \cos\eta J_1(ka\sin\eta) \mathrm{e}^{\mathrm{i} kL/2 \cos\eta} P_l^0(\cos\eta).
    \label{eq:finitetransducer}
\end{equation}

This is therefore the beam shape coefficient for the standing wave formed by the two counter-propagating pressure fields from a finite transducer source.

\subsubsection{Acoustic Force}
As the incident sound waves propagate through the medium, as described by Eq.~\ref{eq:Pinc}, and encounters a spherical bubble, the pressure field is scattered by the bubble and can be described as,

\begin{equation}
     P_{\mathrm{sca}}(r,\theta,\phi)=P_0
     \sum_{l=0}^{l_{\mathrm{max}}} \sum_{m=-l}^{l}C_{lm}h_l^{(1)}(k^E r) Y_{lm}(\theta,\phi),
 \end{equation}

where the scattering coefficient $C_{lm}$, can be determined are determined from two boundary conditions for the acoustic fields around the bubbles. With the pressure and velocity field continuity boundary conditions, one can write the scattered coefficients as \cite{pessoa2020acoustic},
\begin{equation}
    C_{lm} = G_{lm}\left[\frac{\alpha j^{\prime}_l(k^Ea)j_l(k^Ia)-j_l(k^Ea)j^{\prime}_l(k^Ia)}{h_l^{(1)}(k^Ea)j^{\prime}_l(k^Ia) - \alpha h_l^{\prime(1)}(k^Ea)j_l(k^Ia)}\right]=G_{lm}C_l,
 \end{equation}
with $j_l$ and $j^{\prime}_l$ representing the spherical Bessel functions, and derivatives, respectively, and $h_l^{(1)}$ $h_l^{\prime (1)}$ are the spherical Hankel functions of first kind, and their derivatives, respectively. Here, $\alpha \equiv \frac{\rho^Ic^I}{\rho^Ec^E}$, where $\rho$ and $c$ indicate the density and sound speed in the internal and external media. Having computed the scattering coefficient, it is possible to calculate the acoustic radiation force by integrating the incident and scattered pressure fields over a sphere placed at infinity. This is expressed by,
\small
\begin{equation}
    \boldsymbol{F}_{\mathrm{acoustic}}(r)= -\int \,dA \bigg \{\bigg[ \kappa \frac{1}{2} \langle P_t^2 \rangle - \frac{1}{2} \rho^E \langle v_t^2 \rangle \bigg ] \hat{r}+\rho^E \langle(\hat{r}\cdot \vec{v_t})\rangle \bigg \},
    \label{eqfrad}
\end{equation}
\normalsize
where $\vec{v_t}$ is the field velocity defined as $\boldsymbol{v} = -\frac{i}{\rho \omega} \nabla P$ solved in spherical coordinates,  $\kappa=1/\rho^E c_E^{2}$ is relative to the external medium (sunflower oil, in this case) with velocity $c_E$ and the brackets $<.>$ represent a time-averaged quantity. The term $P_t$ is a sum of the incident and scattered beam coefficients, $G_{lm}$ and $C_{lm}$, respectively, expanded into spherical harmonics. 

\begin{equation}
\begin{split}
    F_x + \mathrm{i} F_y &= -\frac{F_0}{4} \sum_{l=0}^{l_{\mathrm{max}}} \sum_{m=-l}^{l} \frac{\mathrm{i}}{\sqrt{(2l+1)(2l+3)}}\left[ \sqrt{(l-m+1)(l-m)} Ct_l G_{l,m+1}G_{l+1,m}^{*} +\right. \\ 
    &\left.+\sqrt{(l+m+2)(l+m+1)}Ct_l^{*}G_{l+1,m+1}G_{l,m}^{*} \right] ,
    \label{fradx}
\end{split}
\end{equation}

\begin{equation}
    F_z=F_0 \operatorname{Im}\left[\sum_{l=0}^{l_{\mathrm{max}}} \sum_{m=-l}^{l}\sqrt{\frac{(l-m+1)(l+m+1)}{(2l+1)(2l+3)}} Ct_l G_{lm} G_{l+1,m}^{*} \right],
    \label{fradz}
\end{equation}

where $Ct_l=\left( 2 C_l C_{l+1}^* + C_l+C_{l+1}^*\right)$, and a force amplitude of $F_0=|P_0|^2/\rho \omega^2$, where $\omega=k^Ec^E$. Note that one might be tempted to simplify these expressions, given our azimuthal symmetry which leads to $m=0$. But, this is only for the special case where the scatterer is located at the origin of our finite stationary wave field. To treat the case where the scatterer can be positioned arbitrarily in space ($\rho_0,\phi_0,z_0$) with respect to the beam coordinate system, we displace the beam and maintain the scatterer at the origin instead, similar to what has been previously showed in Ref.~\cite{pessoa2020acoustic}. To achieve this, we rewrite the original beam coordinates in the following coordinate transform,
\begin{equation}
    \begin{split}
        x&=r\sin\theta \cos \phi - \rho_0 \cos \phi_0 \\
        y&=r\sin \theta \sin \phi - \rho_0 \sin \phi_0\\
        z&= r \cos \theta - z_0.
    \end{split}
\end{equation}

Instead of the coordinate transform for the non--shifted origin presented just before Eq.~\ref{eq:fullPfield}. With this transform we obtain, following the same procedure, the beam shape coefficient as a function of the scatterer position (i.e. origin of the coordinate system), as  
\begin{equation}
\begin{split}
G_{lm}^{SW}&=8\pi \mathrm{e}^{-i m \phi_0} (ka) \sqrt{\frac{2l+1}{4\pi}\frac{(l-m)!}{(l+m)!}}\int d\eta \,\cos\eta \mathrm{e}^{\mathrm{i} k L/2 \cos\eta} \\
    & \cos\left(k z_0\cos\eta-\frac{(l-m)\pi}{2}\right)  J_1(ka\sin\eta) J_m(k \rho_0\sin\eta) P_l^m(\cos\eta).
    \end{split}
\end{equation}

Using this beam shape coefficient in Eqs.~\ref{fradx} and ~\ref{fradz}, the acoustic force profile can be determined in 3-D for the standing wave. 
 
\section{Results and discussion}\label{sec:results}
\subsection{Bubble Tracking}

\begin{figure}[h!]
\centering
\includegraphics[height=4.5 cm]{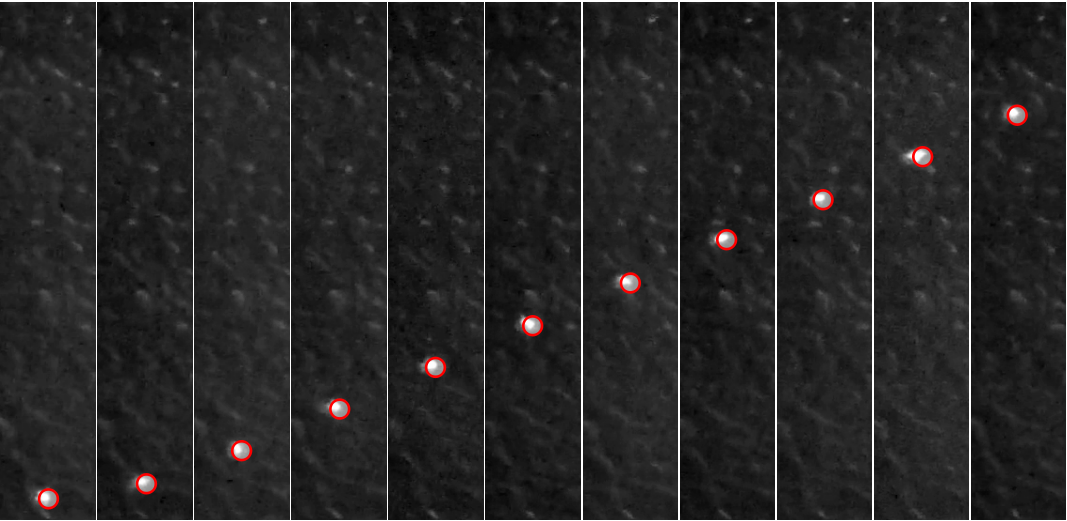}\,\includegraphics[height=4.5 cm]{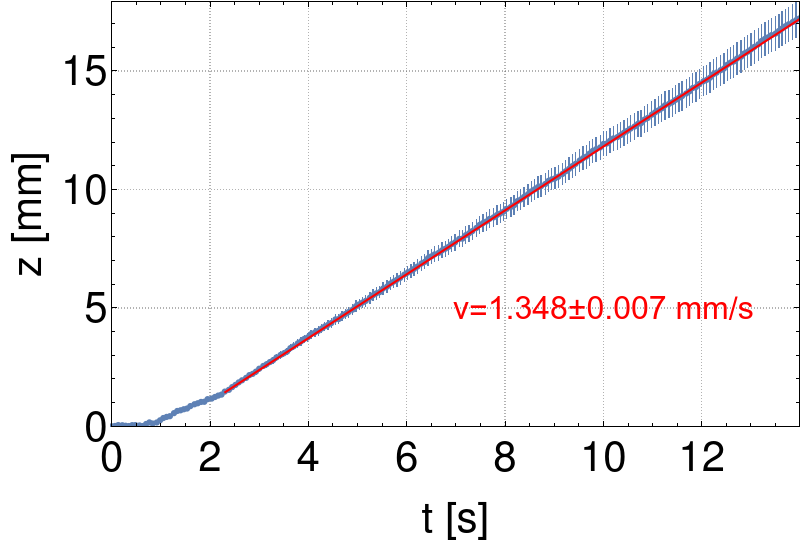}
\caption{(left) Sequence of tracking results for a free bubble. (Right) Displacement tracked for each time step, and linear fit (red line) resulting in the indicated speed and uncertainty.}
\label{dem}
\end{figure}

Bubble tracking was made using Mathematica (Wolfram Research, Inc.), for each experiment video was recorded just before the bubbles where released. For each video frame, a rough estimate for each of the bubble position was determined by an intensity centroid algorithm. From each intensity centroid region, a Gaussian intensity was used for precise sub--pixel tracking (Fig.~\ref{dem}(left)). From the Gaussian fitting, position center and uncertainty was obtained as a function of time. From the precise tracking results, for each bubble, a linear fitting recovered the terminal velocity of the rising air bubble in oil (Fig.~\ref{dem}(right)). The rise velocity $v$ was determined with relative errors of the order of $10^{-3}$. 

\begin{figure}[h!]
\centering
\includegraphics[height=4.5 cm]{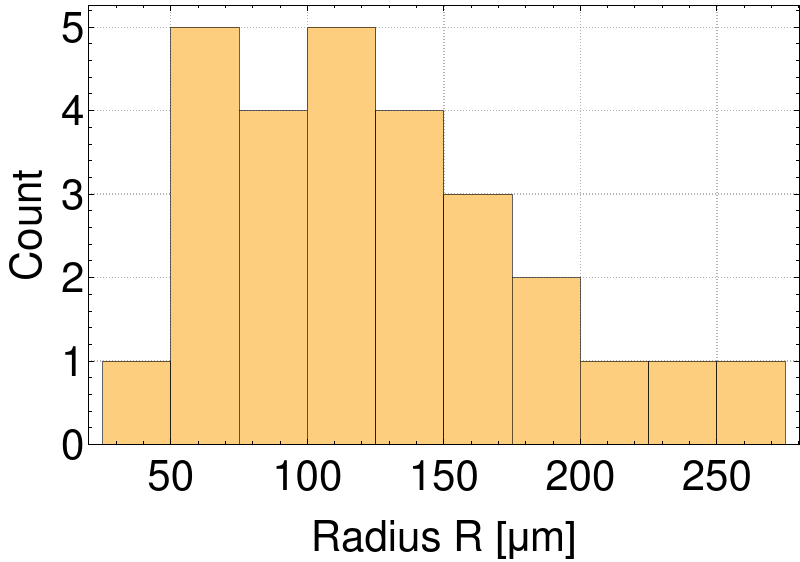}\,\includegraphics[height=4.5 cm]{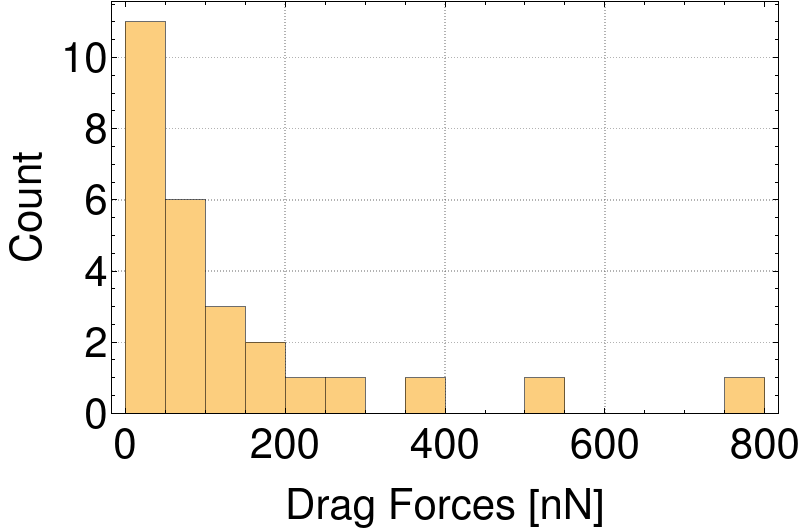}
\caption{Histogram showing the distribution of bubble sizes (left) and corresponding drag forces (right).}
\label{fig:hist}
\end{figure}

Using this experimental data, and the Stokes' drag model, in Eq.~\ref{eq:StokesDrag}, we can determine the bubble radius. The range of bubble radii results are presented in Fig.~\ref{fig:hist}(left) as a histogram considering the 27 bubbles analyzed. Bubbles ranging from $R=43.6$ \textmu m to $R=267.6$ \textmu m were successfully trapped by ultrasonic standing waves. Knowing the bubble size and speed, the drag force can be determined (Fig.~\ref{fig:hist}(right)), which ranges from $3$~nN to $780$ nN. 

Considering the bubbles in a steady position due to the acoustic force $\textbf{F}_{\mathrm{acoustic}}$, then this force must equal to the drag force by the bubble. Using the values obtained for the radius, we can determine the maximum acoustic force, from simulations, and compare it with the drag force.

\subsection{Standing Wave Beam Description}

\begin{figure}[h!]
\centering
\includegraphics[width=0.9\textwidth]{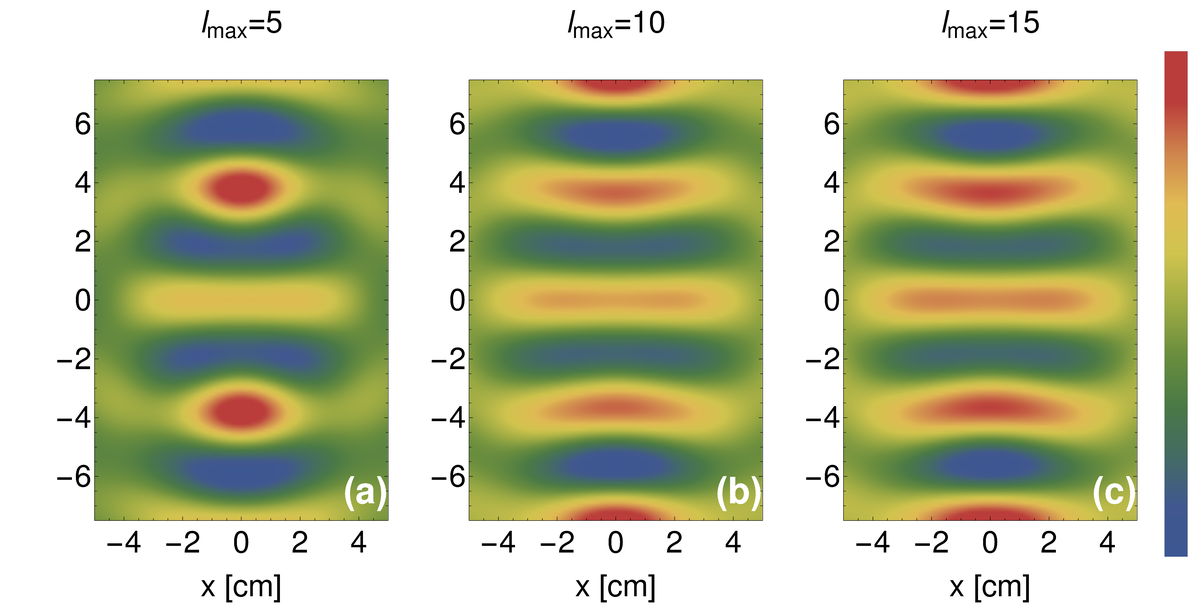}
\caption{Comparison for reconstructed standing wave beam, using GLMT, as a function of the maximum number $l_{\mathrm{max}}$ of terms being summed (a) $l_{\mathrm{max}}=5$, (b) $l_{\mathrm{max}}=10$, and (c) $l_{\mathrm{max}}=15$.}
\label{fig:studyL}
\end{figure}

We performed numerical simulations and integration using Mathematica to visualize the behavior of the pressure field described by the angular spectrum, Eq. \ref{eq:FiniteStatWavePressureField}, and the reconstructed beam shape coefficient using the GLMT framework, Eq. \ref{eq:finitetransducer}. For the reconstructed beam shape coefficient, we performed a study on the dependence with $l_{\mathrm{max}}$, since the sum of infinite terms, must be truncated at a finite value for simulations, shown in Fig. \ref{fig:studyL}. In this figure, we reproduce the experimental scenario where the two piezoelectric transducers are located $L=15$~cm apart and have a radius of $a = 2.5$~cm. The frequency used in the simulations was identical to the experimental, $f = 38.450$~kHz. The piezoelectric transducers are centered at $x=0$, an edges at $x = \pm 2.5$~cm and at $z = \pm 7.5 cm$. In~Fig.\ref{fig:studyL} (a), we see that for $l_{\mathrm{max}}=5$, we see three well--defined pressure zones: the first nodes that come from each transducer, and a zone at the center with slightly lower intensity, but no significant pressure values near the surface of the transducers, which motivates using higher values of $l_{\mathrm{max}}$. In~Fig.\ref{fig:studyL} (b), for $l_{\mathrm{max}}=10$, we see a better--defined pattern in the medium, and at the surface of the transducers, therefore a more realistic field. For $l_{\mathrm{max}}=15$ in~Fig.\ref{fig:studyL}(c), we observe the same pattern as in~Fig.\ref{fig:studyL}(b), but with slightly higher intensities, a sign of convergence. Furthermore, we also investigated $l_{\mathrm{max}}=20$, and $l_{\mathrm{max}}=50$, and observed no significant changes in the pressure field, and therefore we defined $l_{\mathrm{max}}=20$ as adequate for our further investigations. 

\begin{figure}[h!]
\centering
\includegraphics[width=0.9\textwidth]{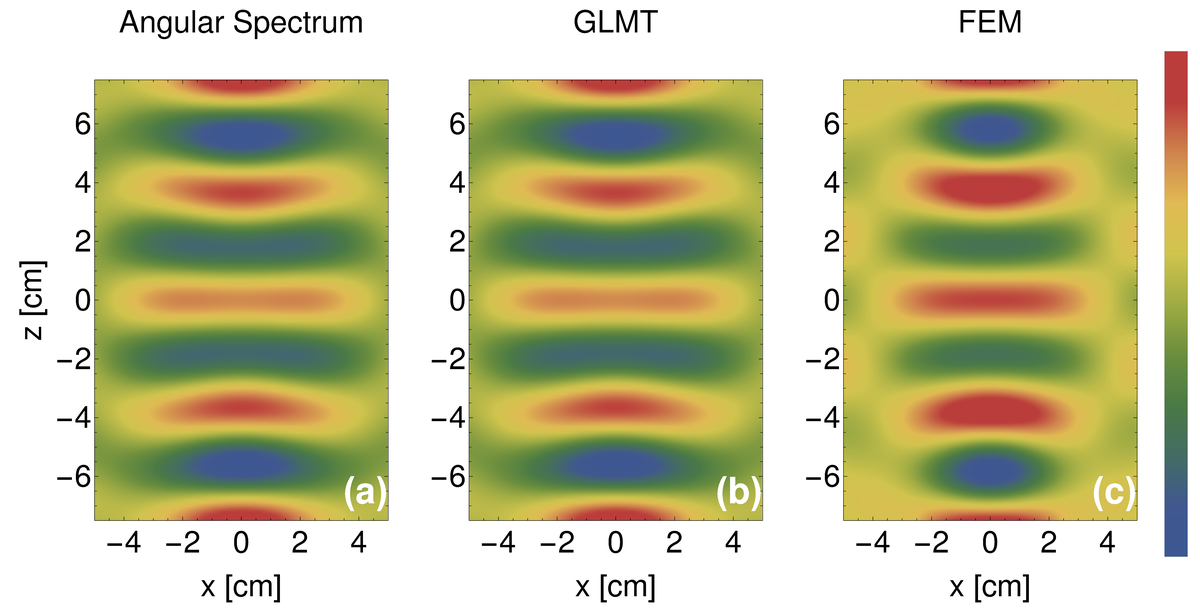}
\caption{Comparison between the stationary wave pressure field from (a) angular spectrum representation, (b) reconstructed from the beam shape coefficients with the GLMT, with $l_{\mathrm{max}} = 20$, and (c) finite elements methods from COMSOL simulations.}
\label{fig:CamposReconstruidos}
\end{figure}

\begin{figure}[h!]
\centering
\includegraphics[width=0.9\textwidth]{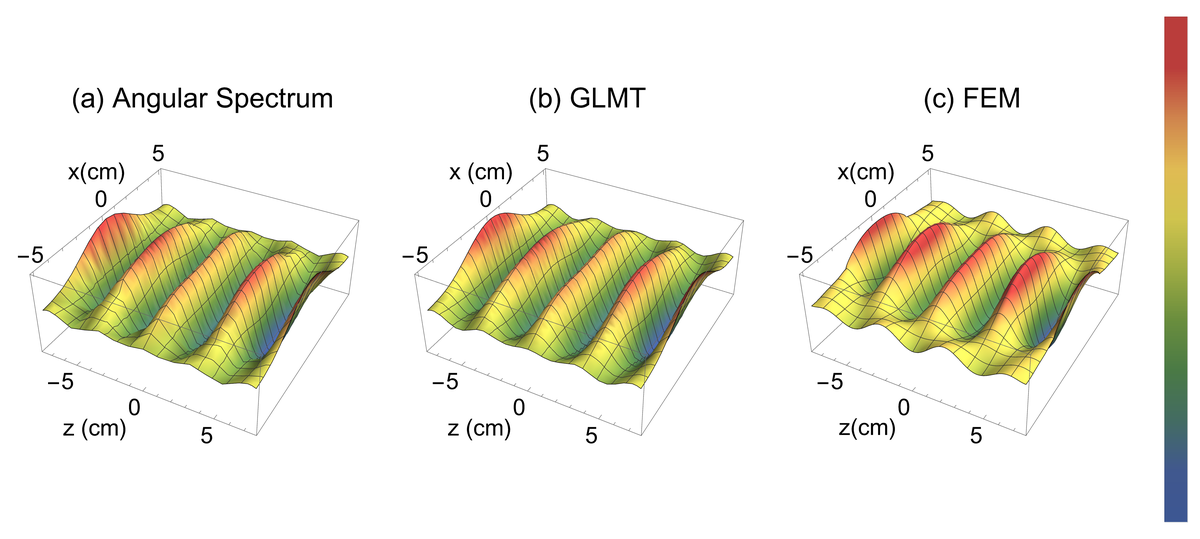}
\caption{Same as Figure \ref{fig:CamposReconstruidos}, but in 3-D. Neither in (a) or (b) we accounted for the boundaries around the recipient, whereas in (c), we did.}
\label{fig:Campos3D}
\end{figure}

In Fig. \ref{fig:CamposReconstruidos}, we present the comparison between the Angular Spectrum representation to describe the pressure field in (a), the GLMT technique in (b), and a simulation using a finite elements method (FEM) in COMSOL Multiphysics (COMSOL). For all cases, we used $f = 38.450$~kHz, $c_E = 1400$m$s^{-1}$, and $\rho_{oil}=988.16$ kg/m$^3$. The COMSOL simulations were made considering piezoelectric transducers made of a lead of zirconate titanate with a radius of $a = 2.5$~cm, being surrounded by a rectangular box of $10$~cm in width and $15$~cm in length. We made use of the stationary solver for wave propagation and interfaced with the multiphysics tool to account for electrostatics and solid mechanics effects at the surfaces of the transducers, and the associated boundary conditions. We implemented extra-fine triangular mesh to prevent convergence issues. 

In Fig. \ref{fig:CamposReconstruidos} (a),(b),(c) the nodes and antinodes are located in the same positions in space, with a width comparable to the size of the transducer, and a distance of a few centimeters in between them.
We notice the diffraction effect of the spherical waves in the $z$ direction, and a curvatures in the nodes due to the superposition of the pressure waves emitted by the transducers. Noticeably, at the center, the interference pattern flattens the pressure field, creating a zone with a higher pressure intensity. For values of $x<-4$~cm and $x >4$~cm the field is practically uniform, without the formation or span of the nodes. This is a particularly important case, as the transducers have their edges placed at $x=(-2.5,2.5)$~cm. 

In \ref{fig:CamposReconstruidos} (c), at $x=(-5,5)$~cm, we placed the boundaries of a rectangular box to match the experimental case where we have a rectangular flask between the two transducers. Even with a larger rectangular box, we observed no relevant changes in the intensity or location of the nodes and antinodes in the central region where the bubbles were trapped. The only observed difference was in the formation of lower intensity nodes which are characteristic of far--field pattern. At a given distance from the source, in this case the transducers, are considered point sources, and the spherical propagation of waves creates those additional nodes. 

In Fig. \ref{fig:Campos3D}, we present a 3-D view of the pressure field. In Fig.  \ref{fig:Campos3D}, at the vicinity of $x=0$~cm, we see that the pressure field follows the cylindrical shape of the transducers that were included in the simulation. The addition of transducers also constrains the pressure field in the $x$ direction comparing the overall shape of the beam (c) to (a) and (b). The differences between the analytical solutions in (a) and (b) and the FEM modeling in (c) can be attributed to the fact that in the FEM COMSOL simulation, the fluid medium is limited by a finite border at $x\pm5$~cm.

\begin{figure}[h!]
\centering
\includegraphics[width=0.8\textwidth]{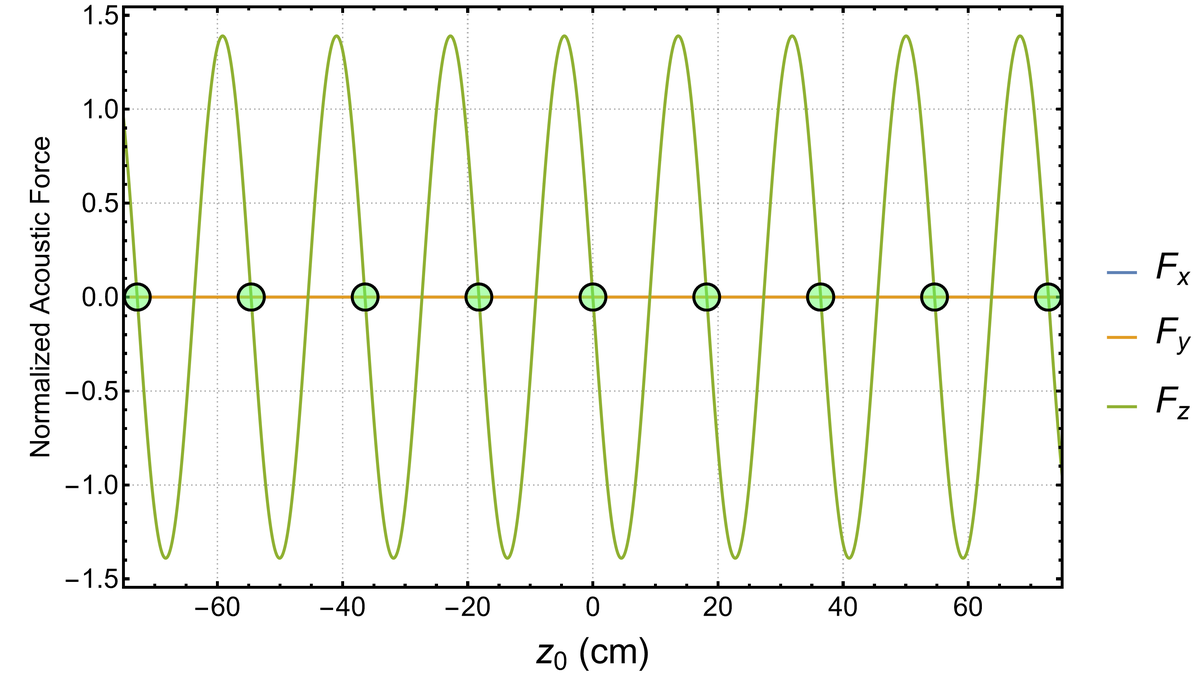}
\caption{Components $F_x$, $F_y$, and $F_z$ of the normalized acoustic radiation force for the infinite plane wave expansion case, acting on a bubble with a typical size of $R=100$ \textmu m. The green dots represent regions where the bubbles can be successfully trapped in the $z$ direction.}
\label{fig:acousticforceplane}
\end{figure}

In the case for transducers producing counter-propagating infinite plane waves in oil, the beam shape coefficient may be written as \cite{pessoa2020acoustic},
\begin{equation}
    G^{\mathrm{sta}}_{lm}=2 \sqrt{4\pi(2l+1)} \delta_{m,0} \operatorname{Re}\left[\mathrm{i}^l \mathrm{e}^{-\mathrm{i}k^E z_0} \right],
    \label{eq:gest}
\end{equation}
where $z_0$ represents the longitudinal direction of propagation. The waves coming from both transducers generate pressure nodes in space, but however, there is no point of stable equilibrium in the $x-y$ direction that characterizes a force, as can be seen in Fig. \ref{fig:acousticforceplane}, for a bubble of radius $R = 100$ \textmu m. This means that multiple trapping points periodically spaced in the $z$-direction are created, and for each trapping point, for example $z=z_0$, there is a restoring force which can be determined by $\textbf{F} = - \nabla \textbf{U}$ that pulls the bubble back to a stable equilibrium position. The green disks in Fig. \ref{fig:acousticforceplane} represent the other nodes where the bubbles can be trapped in the $z$-direction. 

In the experimental case for the acoustic trap however, we observe that the bubbles are trapped in all three dimensions, which indicates points of stable equilibrium also in the transverse coordinates. The acoustic trap needs to overcome, at least, the buoyancy acting on the bubbles so it can successfully trap them \cite{}. This justifies the need of a more realistic description of the beam, such as the one we did using the angular spectrum technique.

\begin{figure}[h!]
\centering
\includegraphics[width=0.45\textwidth]{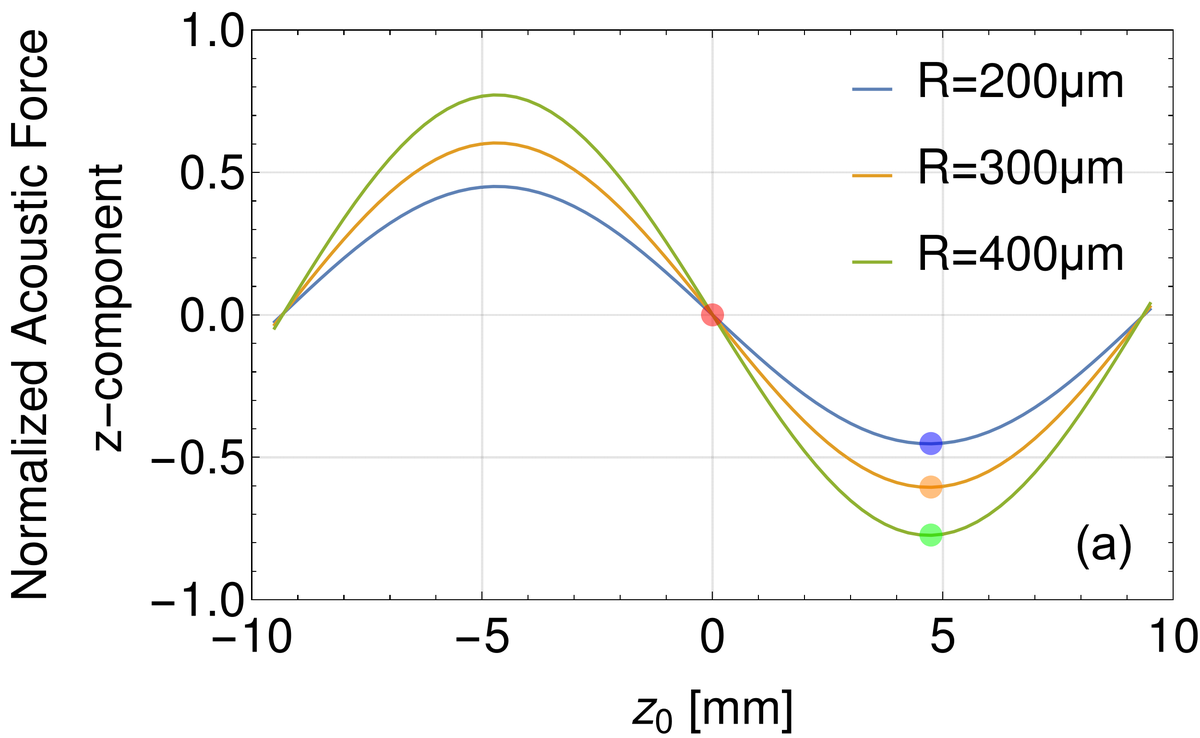}\,\includegraphics[width=0.45\textwidth]{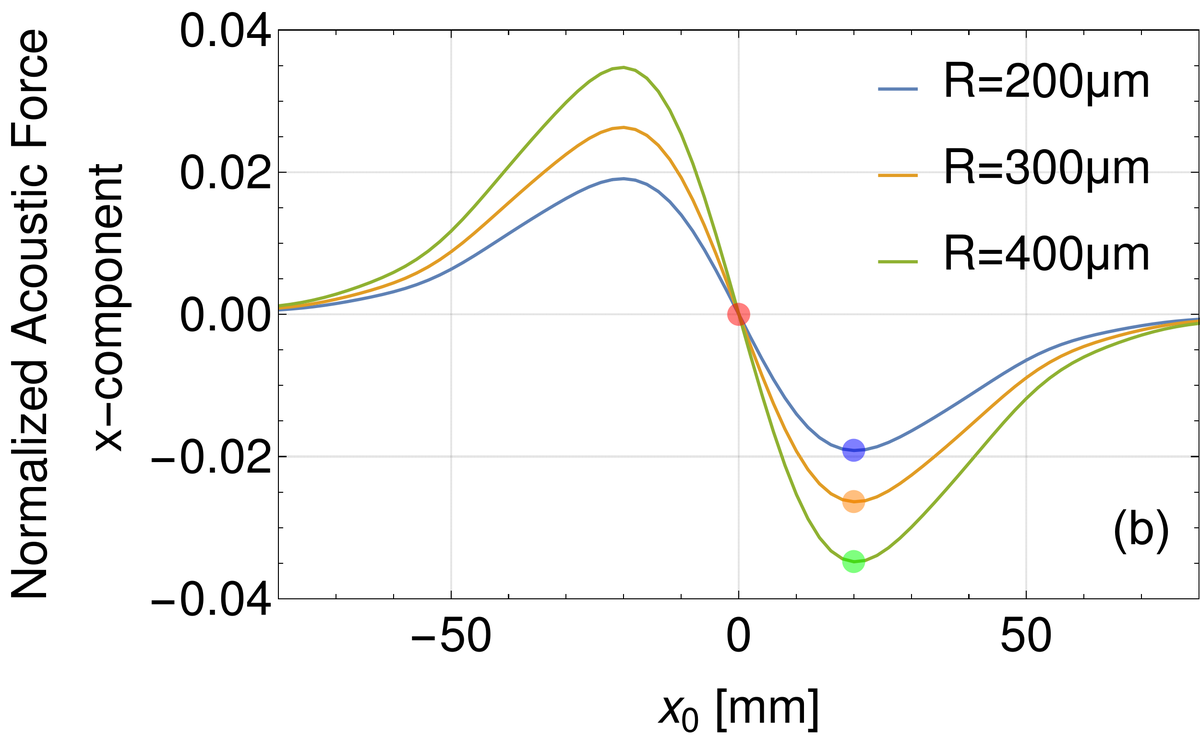}
\caption{Acoustic force components $F_z$ (a) and $F_x$ (b) as a function of distance, for three different sizes of bubbles that were trapped experimentally, $R = 200$ \textmu m in blue, $R= 300$ \textmu m, and $R = 400$ \textmu m. Highlighted with the disks are the stable equilibrium positions.}
\label{fig:finalforce}
\end{figure}

We evaluated the force profile for the standing wave by using the beam shape coefficient in Eqs.~\ref{fradx} and ~\ref{fradz}, for three sizes of bubbles, $R=200$~\textmu m, $R=300$ \textmu m, and $R = 400$ \textmu m. The result is presented in Fig. \ref{fig:finalforce}, in (a) for the $z$ direction, and in (b) for the $x$ direction. The normalized acoustic force is higher in $z$ than $x$, which can be explained by analyzing the fields in Fig. \ref{fig:CamposReconstruidos} and Fig. \ref{fig:Campos3D}. For the pressure field in $z$, noticeably the gradient of pressure, $\nabla P$ is higher than in $x$, and since $F \propto \nabla P$, the force experienced by the bubbles is also higher. The profiles show in (a) and (b) predict that the acoustic radiation force can restore any small displacement around the highlighted equilibrium points in red.

In Fig. \ref{fig:finalforce}(a), we observe, at $z_0=0$, in red, a region where the bubbles can be trapped, where there is the highest potential gradient and the higher pressure intensity, and there is stable equilibrium for the three bubble sizes. The maximum span of the restoring force that can bring the bubbles to a stable equilibrium is $5$~mm, but we see that for each the magnitude changes, as it is expected due to the different sizes. Any physical perturbation on the bubbles up to that point will be counter-acted by the acoustic radiation force. In Fig. \ref{fig:finalforce} (b), the maximum span of the acoustic radiation $x \approx 20$~mm, and we see a trapping point in common with  Fig. \ref{fig:finalforce}(a), at $x_0=0$~mm. The values for the acoustic radiation force that allows the trapping at those positions need to be at least equal in magnitude to the buoyancy force corresponding to each bubble.

The results shown in Fig. \ref{fig:finalforce} are qualitatively similar to the ones described in the literature \cite{wu1991acoustical}, considering the propagation in the $z$ direction, and the behavior of the pressure field creating zones where intensity gradients give rise to an acoustic radiation force.

\section{Conclusion}\label{sec:conclusion}

\quad In this work, we built an acoustic trap for microbubbles, and developed a theoretical analytical model to explain the physics behind the trapping that was further compared to FEM simulations. We used Stokes' drag law to obtain the radius of the successfully trapped bubbles, of $43$ \textmu m up to $275$ \textmu m, and measured the ascension velocity. Knowledge of the drag force and ascending velocity allowed us to estimate the lower bound of the acoustic radiation force, which ranged from $3$~nN to $780$~nN. 

A simple model that uses infinite expansion of counter-propagating plane waves cannot explain why the bubbles could be trapped, since there are no components in the $x-y$ direction. The acoustic radiation force generated by standing plane waves can only be exerted in the $z$ direction, a result that is also in agreement with the velocity and pressure potential theoretical description of this system \cite{bruus2012acoustofluidics}. This motivated the implementation of the angular spectrum technique, widely used in optics \cite{oh2015cylindrical,kozacki2015angular}, to treat this problem using a finite expansion of waves. By using this angular spectrum technique, we were able to find a more realistic beam shape coefficient for the transducer, with which we could find the pressure field using GLMT theory, and calculate the non-zero forces in the $x-y$ direction. The behavior of the pressure field in space was validated by comparing both the angular spectrum beam and the reconstructed by GLMT beam with a FEM simulation in COMSOL. The COMSOL simulation takes into account mechanical properties of the system that were not taken into account in our analytical modelling. 

Thermal and viscous effects, as well as the role of surface tension in bubble trapping \cite{doinikov1998acoustic,lee1993acoustic}, are not taken into account by our formulation of the problem, since our focus was on the beam description. Using the angular spectrum technique, however, it is possible to explain the acoustic trapping in three dimensions by considering solely the characteristics of the beam, and a plane wave expansion within a finite region in space. We show that the beam can be responsible for generating pressure patterns that give rise to an acoustic radiation force for different sizes of bubbles. The calculations presented in this paper can be used to find the acoustic radiation force anywhere in space, as it was a generalized theoretical development using the GLMT technique. 

The angular spectrum finite description of the pressure field allows a more realistic view the physics behind the acoustic trap, but of course, it does not account for all experimental features. This is due to a series of factors that might alter the emission of waves by the transducers, such as impedance matching, optimal operation frequencies, and energy loss through the contact between the transducers and the walls of the recipient, among others. Future analytical models could consider the influence the walls of the recipient in the propagation of waves, and the energy loss due to shear waves generated by the contact with the piezoelectric transducers with the walls.

% FIM DO CONTEÚDO DO DOCUMENTO
% -----------------------------
\bibliographystyle{unsrt} % Estilo de Bibliografia
\bibliography{referencias} % Lista de Referências Bibliográficas

\end{document}